\documentclass[10pt,aps,prb,twocolumn,groupedaddress]{revtex4-1}
\usepackage{xcolor}
\newif\ifproofread

\usepackage{graphicx}
\usepackage{dcolumn}
\usepackage{bm}

\usepackage[nopar]{lipsum}  
\begin{document}

\preprint{APS/123-QED}

\title{Coherent Transport in Y-Junction Graphene Waveguide}

\author{Vahid Mosallanejad}
\email{vahid@ustc.edu.cn}
\author{Kuei-Lin Chiu}%
\author{Guo-ping Guo}
\email{ gpguo@ustc.edu.cn}
\affiliation{CAS Key Laboratory of Quantum Information, and Synergetic Innovation Center of Quantum Information and Quantum Physics, University of Science and Technology of China,Chinese Academy of Sciences, Hefei 230026,China}
\noaffiliation




\date{\today}

\begin{abstract}
We performed a series of theoretical transport studies on Y-branch electron waveguides which are embedded in mid-size armchair graphene nanoribbons (AGNRs). Non-equilibrium Green’s function (NEGF) with different approximations of tight-binding (TB) Hamiltonian has been employed. Using the first nearest hopping approximation, we observed very pronounced conductance quantization, the structure of which depends on geometrical design and shows a spacing of $4e^2/h$, indicating the existence of valley degree of freedom. Moreover, by incorporating the third nearest approximation, we observed seminal plateaus deviated from multiples of $4e^2/h$ conductance, suggesting the lift of valley degeneracy. Finally, Quasi-one dimensional band structure calculations have been performed to study the availability of energy channels and the role of the major geometrical parameters on the transport. 
\end{abstract}

\maketitle


\section{\label{sec:level1} Introduction}

In the past decade, many theoretical studies have been proposed to realize the coherent transports in ideal straight graphene nanoribbons (GNRs)~\cite{peres2006conductance,gunlycke2007semiconducting,dubois2009electronic}. However, the quantization of conductance, which is the hallmark of coherent transport in 1D systems, is usually missing in $O_2$ plasma etched GNRs due to the edge disorders that produce strong scattering~\cite{lin2008electrical,lian2010quantum, tombros2011quantized,wurm2009interfaces, mucciolo2009conductance,lherbier2008transport,orlof2013effect,baldwin2016effect}. Perhaps this major problem has hindered graphene's application in sophisticated nanoelectronics (such as spin qubit devices) despite its excellent conductivity~\cite{recher2010quantum,chiu2017single,deng2015charge}. An alternative approach to replicate a GNR is to introduce a well-like electrostatic confinement in graphene, also known as graphene waveguide (GWG), in which the crystal structure between side-barriers and quantum well remains intact~\cite{tudorovskiy2007spatially}. As a result of the electrostatic confinement, there are discrete bound states spatially extended along the graphene waveguide. Quantized conductance in 1D systems had long been realized in modulation-doped GaAs/AlGaAs heterostructures where a pair of negatively biased split gates deplete the two-dimensional electron gas (2DEG) underneath and define a 1D channel~\cite{yacoby1996nonuniversal}. The situation for graphene-based waveguide is expected to be different from the aforementioned 2DEG case, since the existence of transparent Klein tunneling could play an essential role and limit the ability of GWG for efficient guiding~\cite{katsnelson2006chiral, allain2011klein}. Interestingly, some theoretical approaches using Dirac equation have in fact suggested that carriers in an infinite GWG is behaving in many respects like the optical waves in an optical-fiber (that is, they display optics-like features)~\cite{beenakker2009quantum, hartmann2014quasi,dragoman2010polarization}. However, it is worth noting that the theory of Klein tunneling on graphene is based on first-nearest approximation (Dirac Hamiltonian) and the assumption of the plane wave solution. A precise but relatively expensive numerical method incorporating few more nearest-neighbors hopping terms, called the Non-equilibrium Green's function (NEGF), can be employed to have a better understanding about the feasibility of the perfect transmission in GWG~\cite{datta2005quantum}. Such a numerical simulation can also provide more details on the conditions that lead to a coherent or imperfect transmission. Current experimental evidences are not, however, in favor of the total reflection of carriers from borders of a GWG as Snell's law would hold in the optical-fiber. In particular, the reported guiding efficiencies do not exceed 80\% and differ among different 1D energy modes~\cite{williams2011gate,rickhaus2015guiding}. In our previous calculations, we have shown that a single GWG (S-GWG) can possess the characteristics of a coherent transport and the desirable insensitivity to the bending degree~\cite{mosallanejad2018perfectly}. A further interesting question for us is whether a coherent splitting of current can be realized in GWGs. In this letter, we provide theoretical  evidences to show that it is possible to coherently split the incoming current into two paths in a Y-junction graphene waveguide (YJ-GWG). 
Our results show the prospect of using YJ-GWGs as a splitter in all graphene-based electronic devices that require robust phase coherence, and therefore can be potentially used as an interconnect in spintronic devices. A YJ-GWG can be made by applying a positive potential to an external Y-branch metal gate which is isolated from graphene flake and consequently it induces a Y-shape potential well in graphene. Mathematically, a reasonable estimation of 2D on-site potential (a cut-plan of 3D electrostatic potential) suitable for modeling of YJ-GWG could be produce by making use of two bended S-GWG with an overlapping area as it explained in appendix A.
The influence of geometrical design on the coherent transport is discussed in detail in the main text. Furthermore, Quasi-one dimensional bandstructure calculations have been performed to study the energy channels that could contribute to the transport.
\section{Methods}
\subsection{Device structure}
\begin{figure*}
\begin{center}
\includegraphics[width=18cm]{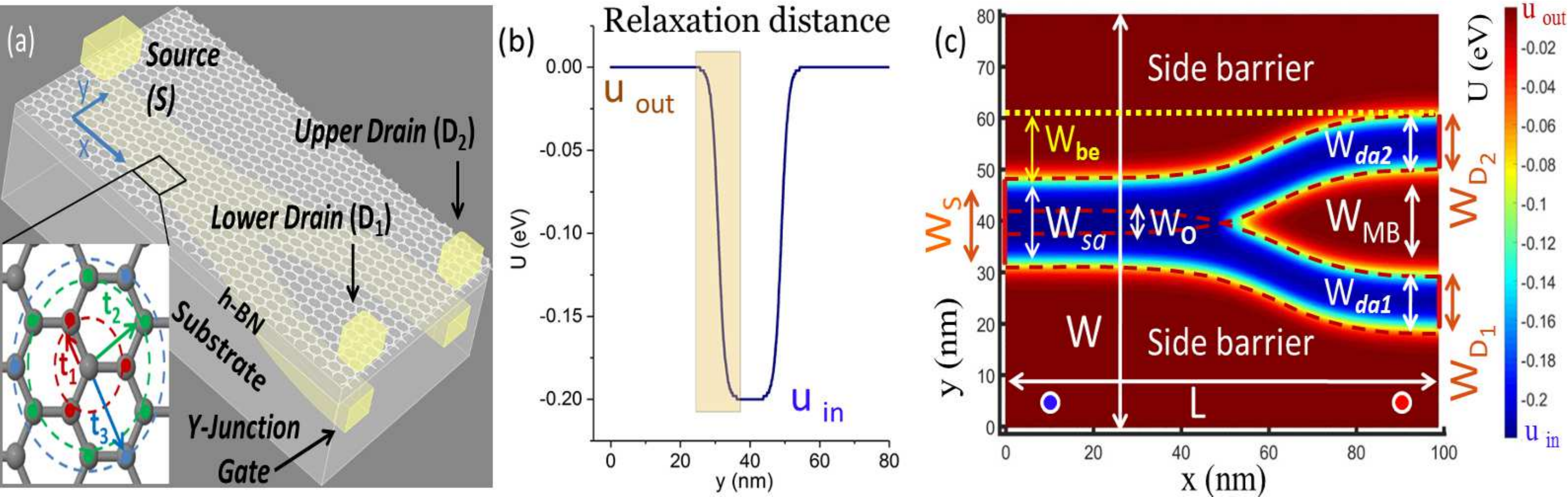}
 \end{center}
\caption{\label{fig1} (a) Schematic illustration of YJ-GWG where a Y-junction metal gate is embedded in an insulator and a mid-size GNR lies above. Enlarged view shows hopping parameters. The crystal structure of graphene is scaled for illustrative purposes and do not represent the realistic crystal size.(b)  (c) Realistic on-site potential of the proposed structure in Fig. 1(a). Source and drains are shown by vertical (red) lines and their widths are denoted by W$_{S}$ and W$_{D_{1,2}}$.}  
\end{figure*}
The schematic of our proposed structure is shown in Fig.~\ref{fig1}(a), which consists of a finite mid-size AGNR with an integrated Y-junction gate underneath it. Hexagonal boron nitride (h-BN), an atomically thin insulating material, is used simultaneously as a gate dielectric and a secondary substrate to improve the mobility of graphene~\cite{zomer2011transfer,zhang2017electrotunable}.
A YJ-GWG can be defined by applying a positive potential to the underlying gate and consequently induces a Y-shape potential well in the AGNR. Single potential well on the left side thus splits into two wells, one goes upward and the other goes downward beyond the intersection, i.e., the system has in total three terminals. 
The source and drain terminals are located exactly above the respective branches of the Y-junction, which is denoted by S, D$_1$ and D$_2$, respectively.
In general, a self-consistent solution of Schr\"{o}dinger equation (in the form of Green’s function formalism) coupling to a Poissons equation should be used to simulate the electrostatics of the device and the transmission through terminals, simultaneously~\cite{akhavan2012phonon, banadaki2015investigation}.
However, this model can not be realized here due to the numerical complexity arising from the large number of atoms in our system.
To compensate the lack of full self-consistent solution, a representative potential well with a cross section as illustrated in Fig.~\ref{fig1}(b) has been used to define YJ-GWG in our AGNR. In a real device, the relaxation distance between the top and bottom of the potential well depends on the thickness and dielectric constant of the substrates. The thinner the insulators (e.g., hBN) are, the sharper the potential well will be. The on-site potential of carbon atoms in a real 2D device is shown in Fig.~\ref{fig1}(c), in which the YJ-GWG formed by the combination of two S-shape waveguides, is indicated by dashed lines.
The 2D device (AGNR) consists of two main parts. First, a fairly large intrinsic channel with width W and Length L, while $W_{sa,da1,da2}$ represent the width of three arms of YJ-GWG, i.e., \textit{source arm}, \textit{lower drain arm} and \textit{upper drain arm}, respectively.  
In the standard convention, AGNRs are labeled as N$_A$-AGNR where N$_A$ is the number of dimer lines defined by N$_A=\lfloor{W/(0.5\sqrt{3}a_{cc})}\rfloor$, W is the width of AGNR and $a_{cc}=0.142$~nm is the carbon-carbon bond length.
The second part of our device is the contacts (source and drains) which are also made of carbon and are in fact extensions from the scattering area, as illustrated by vertical red lines in Fig.~\ref{fig1}(c).
The widths of all terminals are tailored to be matched with waveguide arms, i.e., W$_{S}$ = W$_{sa}$, W$_{D_{1,2}}$ = W$_{da_{1,2}}$.  
In all numerical examples, we have chosen metallic type of supercells (or (3m+2) family of AGNRs) for all terminals to artificially replicate a real metal contact.   
The reason to use these GNRs instead of metal is to provide good ohmic contacts owing to the workfunction match.
The length of scattering area (L) is related to the number of  supercells (N$_S$) via N$_S=\lfloor{L/(3a_{cc})}\rfloor$.
W$_O$ represents the amount of overlap between the two flat ends of S-shape GWGs, and $W_{MB}$ is the width of middle barrier between the two $drain~arm$s. W$_{sa}$ and W$_{MB}$  are related to W$_O$ via W$_{sa}$=W$_{da_1}$+W$_{da_2}$-W$_O$  and  W$_O$/2+W$_{MB}$/2=W$_{be}$, where W$_{be}$ can be viewed as the amount of bending. 
\subsection{Numerical model}
The $p_z$ orbital wavefunction of Carbon is used to construct TB-Hamiltonian for channels and terminals,  
\begin{equation}
\label{eq:1}
H= \sum_{\langle {ij} \rangle}- t_{ij} a_i^{\dagger} a_j + \sum_{i} v_i a_i^{\dagger} a_i,
\end{equation}
where $a_i^{\dagger}$ and $a_i$ are the creation and annihilation operators at $i$-th atomic site, respectively. The symbol ${\langle {ij} \rangle}$ denotes a pair of atomic sites and the relevant hopping parameters $t_{ij}$ are real energy values. 
\begin{table}
\begin{center}
\caption{\label{tab:table1}TB parameter sets.}
		\begin{tabular}{cccccccc}
         \hline
         \hline
			Set&$E_{2p}(eV)$&$t_1(eV)$&$t_2(eV)$&$t_3(eV)$&$s_1$&$s_2$&$s_3$\\
			\hline
			1NN& 0 &-2.78 &0  &0 &0&0&0\\
			3NN$_I$&-0.45 &-2.78 &-0.15  &-0.095 &0.117&0.004&0.002\\
			3NN$_{II}$& -0.187 &-2.756 &-0.071  &-0.38 &0.093&0.079&0.070\\
            \hline
            \hline
		\end{tabular}
        \end{center}
\end{table}
We have assumed the first nearest and two other third nearest approximations denoted by 1NN and 3NN$_{I,II}$ in different cases to comprehend the impact of them on transport. The relevant TB parameters for 1NN and 3NN$_{I,II}$ are shown in Table~\ref{tab:table1}. 
In particular, the new TB parameters, 3NN$_{II}$, obtained from density functional theory (DFT), has shown excellent agreement with the predictions of the DFT even in the high-energy region~\cite{reich2002tight, kundu2011tight,tran2017third}.
The effect of the potential well on the transmission is directly considered through the on-site potential energy on the Hamiltonian, i.e., $v_i$ in Eq. (\ref{eq:1}).
The assigned potential map $U$ for the Y-junction is shown in Fig.~\ref{fig1}(c), in which the value of $v_i$ in the bottom of Y-shape potential well is $u_{in}$, and that in the barrier regions has a fix value $u_{out}=0$.
Similarly, on-site potential of source and drains terminals in their Hamiltonian are noted as $u_S$ and $u_{D_{1,2}}$, which are used to represent an applied bias voltage. 
The effect of edge roughness in the GNR is not included in our calculation~\cite{gunlycke2008tight}. Because the edge roughness can be far from the potential-defined Y-junction and if a very wide GNR is considered, their effect on the transport in YJ-GWG can be neglected. 
The effect of electron-phonon interactions is not included in this work either, due to the weak electron-phonon interactions in graphene~\cite{borysenko2010first}.
Using the Landauer-B\"{u}tticker formalism~\cite{datta2005quantum}, the conductance between the source and drains in  low temperature can be expressed as:
\begin{equation}
\label{eq:2}
{\rm G_{SD_{1,2}}(E)}={\rm G_0}{\rm Tr}[{\Gamma_{\rm S}(E)} {G^{\rm r}(E)} {\Gamma_{\rm D_{1,2}}(E)} {G ^{\rm a}(E)}],
\end{equation}
where ${\rm G_0}=2e^2/h$ is the quanta of conductance that includes the spin degree of freedom, $\Gamma_{\rm S,~D_{1,2}}$ are the source and drains broadening matrices, which couple the left and right terminals to the central scattering region in YJ-GWG. The retarded Green's function $G^{\rm r}$ is defined as ${G^{\rm r}(E)}=[(E+i\eta)S-H-\Sigma_{\rm S}(E)-\Sigma_{\rm D_{1,2}}(E) ]^{-1}$ and $G^{\rm a}=(G^{\rm r})^{\dagger}$, where S is the overlap matrix and is in the form of the first part in equation (\ref{eq:1}).
The open boundary condition of terminals is added via $\Gamma_{\rm S,D_{1,2}}(E)=i(\Sigma_{\rm S,D_{1,2}}^{\rm r}(E) -\Sigma_{\rm S,D_{1,2}}^{\rm a}(E)) $, where $\Sigma_{\rm S,D_{1,2}}^{\rm a}=(\Sigma_{\rm S,D_{1,2}}^{\rm r})^{\dagger}$ and $\Sigma_{\rm S,D_{1,2}}^{\rm r}$ are self-energies of the semi-infinite terminals on the source and drains. 
The transport simulation of YJ-GWG  is carried out by our ballistic code in which we have used the direct recursive algorithm to calculate conductance of a system with few hundred thousand of atoms~\cite{lewenkopf2013recursive, thor2014recursive}.
This Memory-friendly approach allows us to perform partial inversion of a gigantic matrix, which is a necessary mathematical operation to calculate the Green's function in real space.
\section{Results and Discussion}
\subsection{Tunning}
Y-junction gate voltage is used to control the depth of quantum well and consequently provide deep-enough confined energy channels. By this manner, the bottom of on-site potential, i.e., $u_{in}$, can be parametrically tuned to achieve a coherent transport in the device. Moreover, our previous work has demonstrated that on-site potential of terminals is an equally vital parameter to achieve the coherent transport in a two-terminal graphene waveguide~\cite{mosallanejad2017perfectly}.
\begin{figure}
\begin{center}
\includegraphics[width=8.75cm]{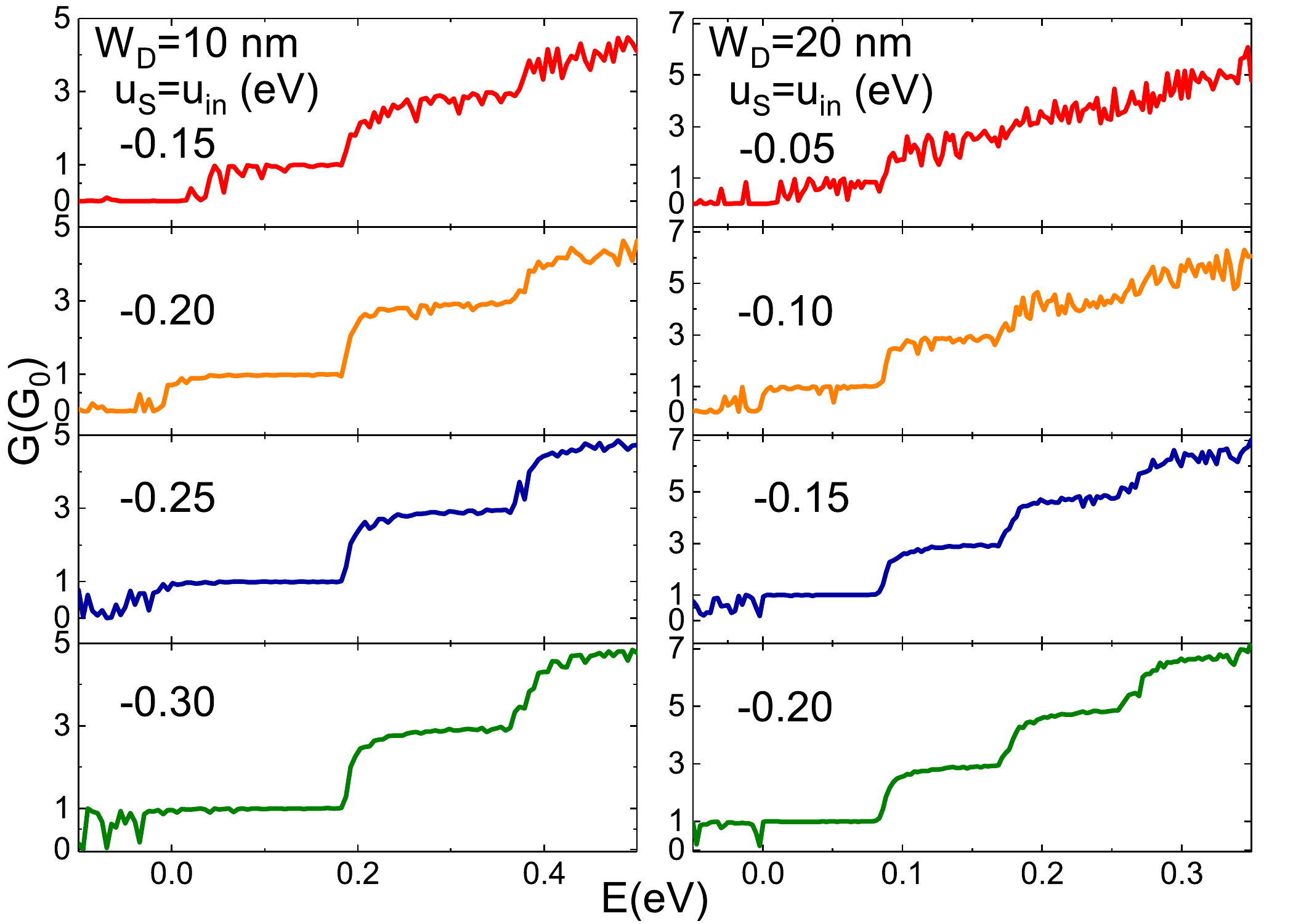}
\end{center}
\caption{\label{fig2} Conductances of two-terminal curved S-GWGs in various $u_{in}$. Right (left) panels show the tunning of a S-GWG with the waveguide width $W_{D}$ of 10~nm (20~nm).} 
\end{figure}
It has been concluded that the quantization of conductance occurs if the following conditions were met: (i) the terminals have similar width of the waveguide, (ii) the source has a fix on-site potential almost equal to that in the bottom of waveguide (i.e., $u_S=u_{in}$), (iii) on-site potential at the drain is kept zero, which is equal to $u_{out}$.To quantitatively determine an appropriate range of $u_{in}$ as a primary study, a few two-terminal transport studies were performed for two curved S-GWGs with waveguide widths ($W_{D}$) of 10~nm and 20~nm.
The on-site potential profile of such systems is similar to that in Fig.~\ref{fig1}(c) but the negative potential well (blue color) is placed only within one arm. For each sample, $u_{in}$ has been swept over a range of values under the conditions of $u_S=u_{in}$ and $u_{D_{1,2}}=0$. Conductances on all upper panels of Fig.~\ref{fig2} (in both sizes), show that shallower wells (small negative $u_{in}$) are incapable to produce smooth plateaus (note that we have employed 1NN approximation in these calculations).
Once the well gets deeper, more well-defined smooth plateaus start to form. Furthermore, narrower waveguide (10~nm) needs deeper potential well to produce smooth plateaus.
Thus, it is essential to have a deep-enough well via applying appropriate gate voltages.
Independent comparison between the right and left panels of Fig.~\ref{fig2} shows that number of plateaus within the same energy range is somehow, insensitive to the depth of potential well but rather dependent on the width of the waveguide. This fact indicates that the width of waveguide determines the conductance profile if the potential is deep enough.
Since $u_{in}=-0.30$ provide smooth plateaus for the 10 nm wide waveguide, hereafter, we will use these quantitative values in the transport study of YJ-GWG in the next section.
\vspace{-1em}
\subsection{Transport study in YJ-GWG}
\begin{figure}
\begin{center}
\includegraphics[width=8.5cm]{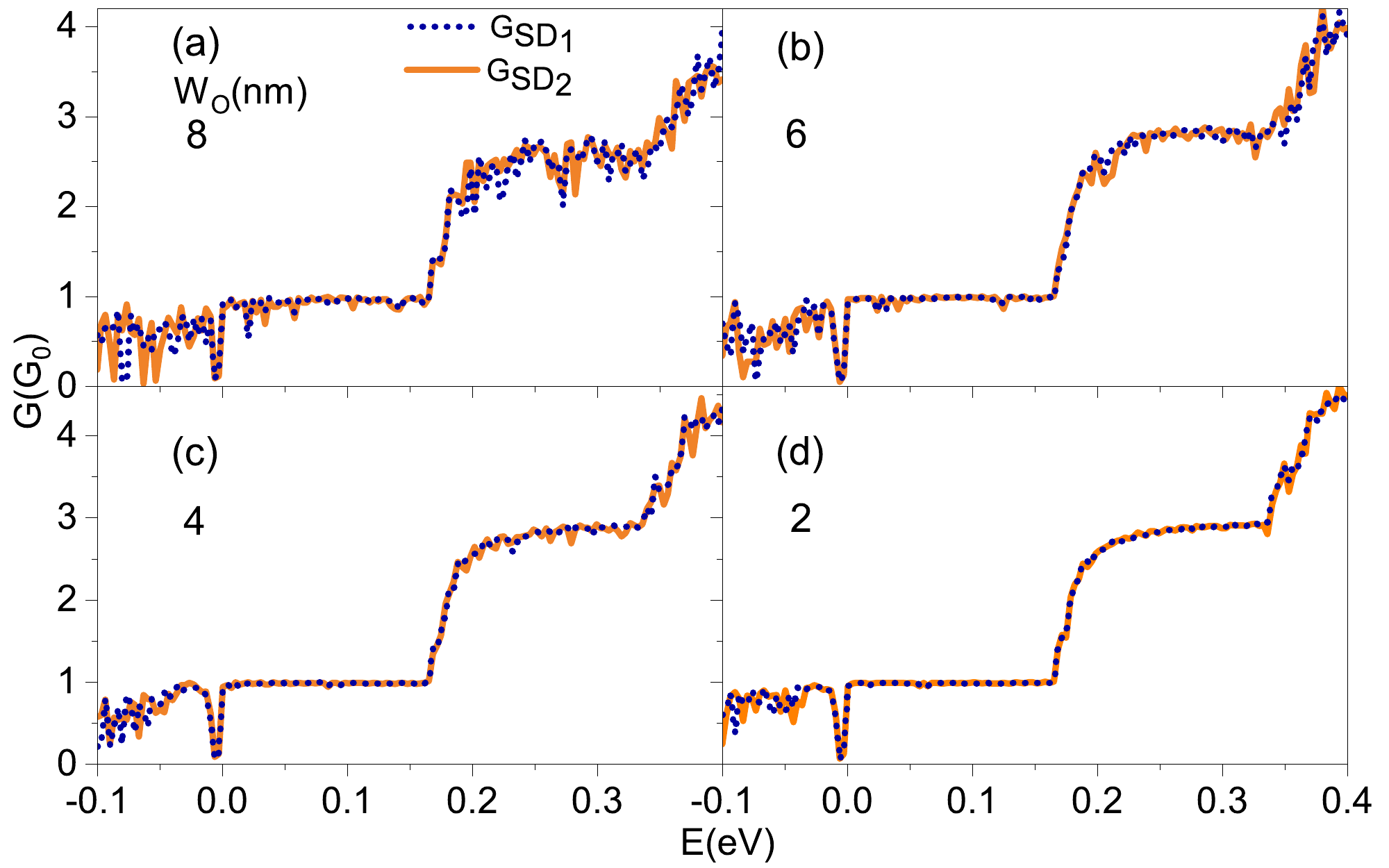}
\end{center}
\caption{\label{fig3} Conductances of both paths in a smooth/S-shape YJ-GWG. Dot and solid lines denote the conductances of lower and upper paths receptively. (a)-(d) are the conductances for different values of $W_O$ as it is noted.}
\end{figure}

\begin{figure*}
\begin{center}
\includegraphics[width=13cm]{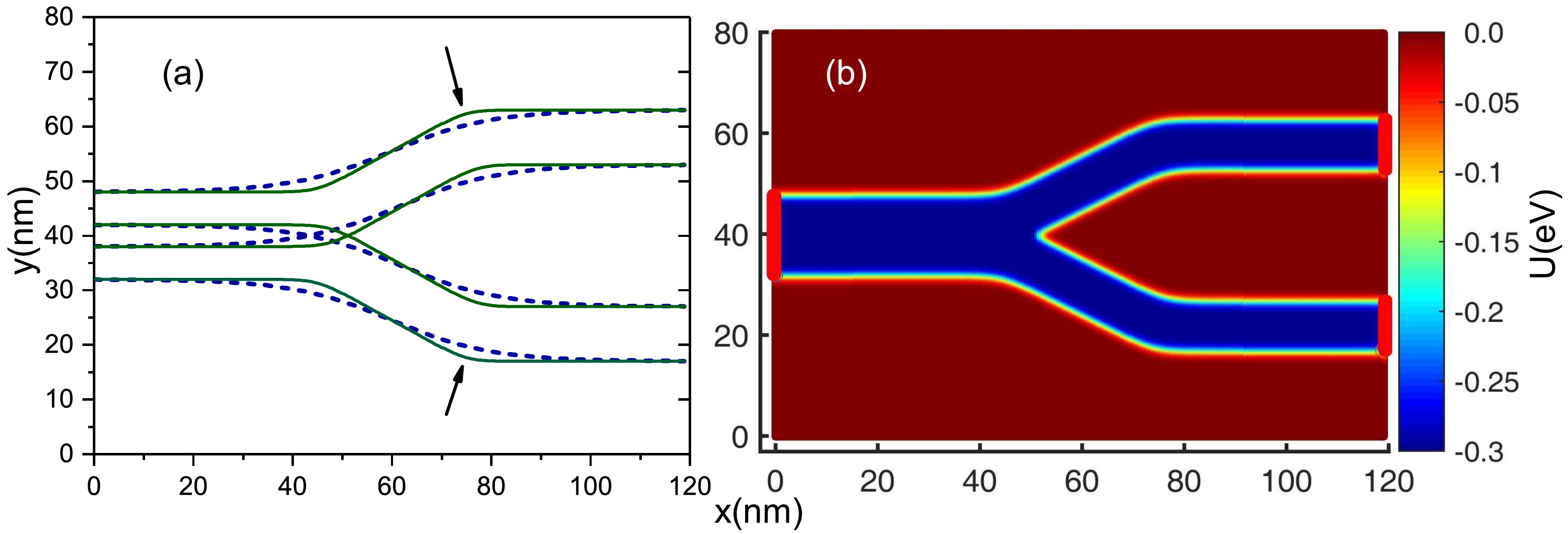}
\end{center}
\caption{\label{fig4} (a) Borders with which potential well of the smooth and non-smooth YJ-GWG have been defined. Dashed lines refer to smooth or S-shape bending whereas the solid lines refer to non-smooth or kink-shape bending. (b) Equivalent potential profile of the non-smooth Y-junction.}
\end{figure*}

\begin{figure}
\begin{center}
\includegraphics[width=8.5cm]{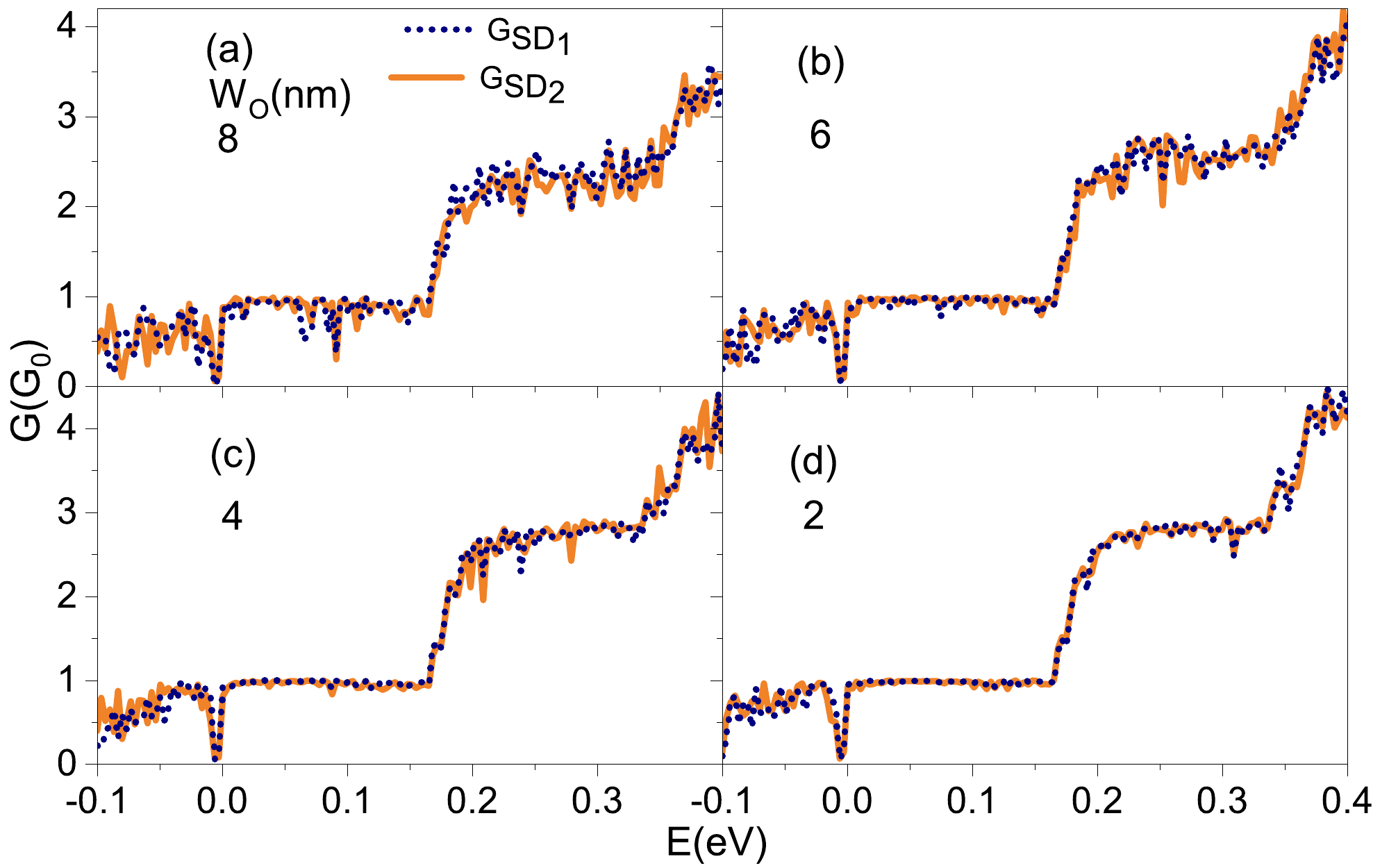}
\end{center}
\caption{\label{fig5} Conductances of both paths in a non-smooth Y-junction graphene waveguide. Dot and solid lines denote the conductances of lower and upper paths receptively. (a)-(d) are the conductances for different values of $W_O$ as it is noted.}
 \end{figure}

 Start with a given drain width $W_{D_{1},D_{2}}=10$~nm, four samples with different $W_O$ and $W_S$ are created and the conductances of both paths in each sample are evaluated.
The dimension of the system is $W \times L=80\times120~nm^2$ and paths have symmetry around $y=W/2$. Center of bending is located in the middle of x-direction at $L/2$. Calculated results are shown in Fig.~\ref{fig3}(a)-(d).
$W_O$ has been selected as the first control parameter to study the effect of overlap, partially because other parameters, such as $W_{be}$, (which represents the amount of bending) has shown very less influences on the conductance of a S-GWG.
Lower panels of  Fig.~\ref{fig3}, (c) and (d), with smooth and identical plateaus for both paths, demonstrate that it is possible to transport carriers coherently in both paths of a YJ-GWG provided that the parameter $W_O$ is small enough.
By comparing all panels in Fig.~\ref{fig3}, one can find a noticeable smooth conductance which was achieved when the width of the source arm increases (because of W$_{sa}$=W$_{da_1}$+W$_{da_2}$-W$_O$).
In addition, another type of YJ-GWG has been constructed based on a kink-shape bending to study the effect of sharp bending. 
Variation of waveguide width occurs around the sharp bending, indicated by arrows in the Fig.~\ref{fig4}(a), within which the width of waveguide differs from $W_{da1,da2}$ in comparison with the fully-smooth S-shape bending shown by dashed lines in the Fig.~\ref{fig4}(a).
S-shape bending ensures that the width of both paths remains unchanged after the splitting point. 
Since the energy channels are highly sensitive to the width of single-waveguide, width variations in sharp bendings are expected to induce extra scatterings that affect the conductance quality.
Fig.~\ref{fig5}(a)-(d) show the calculated conductances of the kink-shape YJ-GWG with different $W_{O}$. Fig.~\ref{fig5} supports our argument and indeed shows more noisy conductances, especially in the second plateau.
Note that except for the shape of the waveguide, all other parameters used in Fig.~\ref{fig5} are similar to those used in Fig.~\ref{fig3} (note that the third nearest approximation, denoted by 3NN$_I$ has been employed for both cases). 

We further have investigated the geometrical effect on transport by enlarging the whole device by a scale factor of 1.5 but keeping the Y-shape structure unchanged. Accordingly, scattering area of $W \times L=120 \times 180~nm^2$, arm's width of $W_{da_{1},da_{2}}=15$~nm, $W_O=3$~nm and $u_S=u_{in}= - 0.25~(eV)$ have been taken into the account.
\begin{figure}
\begin{center}
\includegraphics[width=7.5cm]{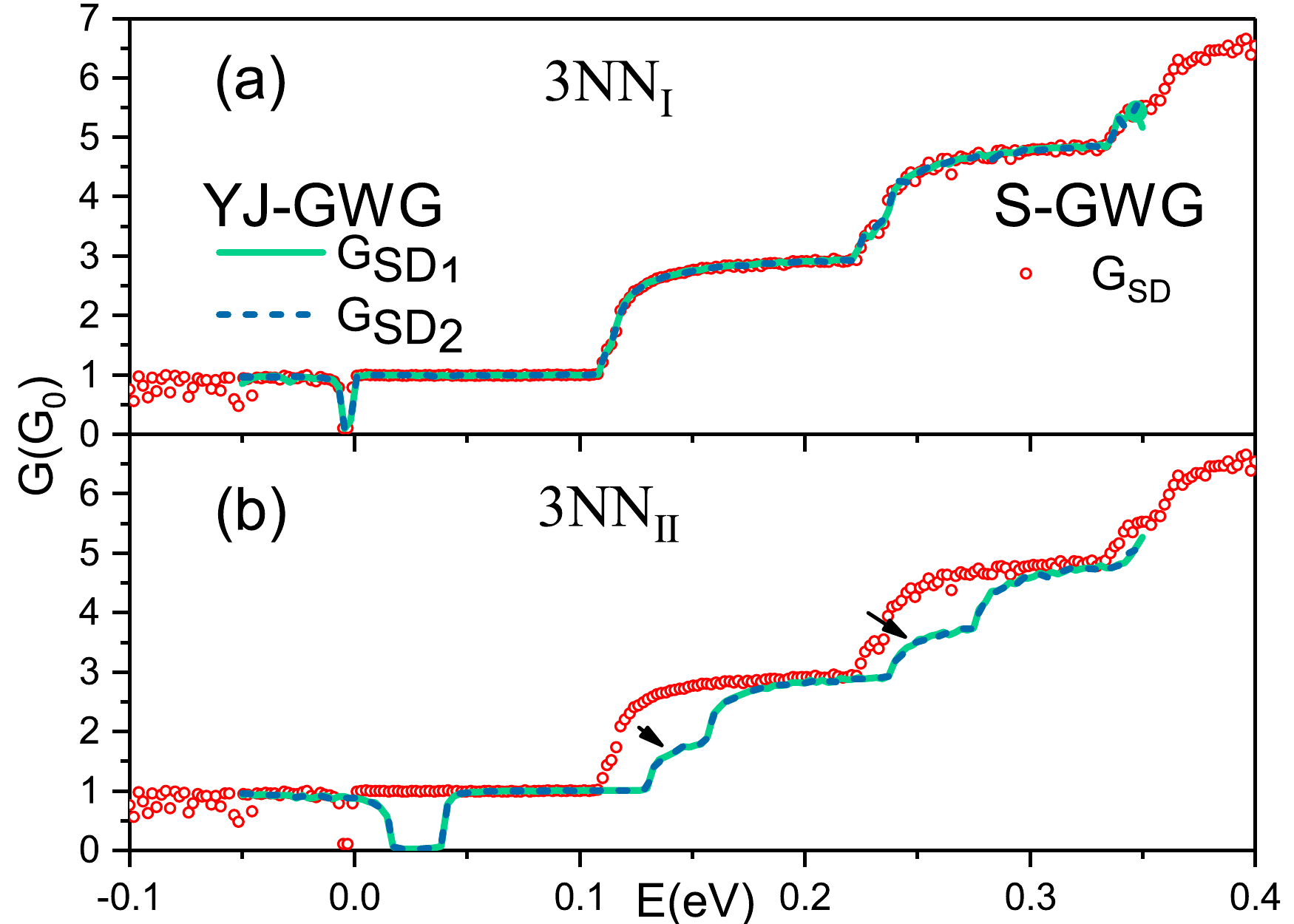}
\end{center}
\caption{\label{fig6} (a) Color lines refer to the conductances of both paths of a YJ-GWG scaled by factor 1.5, employing energy set of 3NN$_I$. (b) Color lines refer to the conductances of same configuration utilizing energy set of 3NN$_{II}$ from Table.~\ref{tab:table1}. Red-hallow circles refers to the two terminal conductance of the relevant S-GWG employing 3NN$_I$.}
\end{figure}
Conductances of two paths have been shown by dashed and solid lines in Fig.~\ref{fig6}(a).
Comparison of conductances between Fig.~\ref{fig3}(d) and Fig.~\ref{fig6}(a) implies that scaling has minor effect on general characterizations of the YJ-GWG.
In Fig.~\ref{fig6}(a), the third nearest approximation, denoted by 3NN$_I$  has been used whereas 3NN$_{II}$ is employed to calculate conductances of YJ-GWG in Fig.~\ref{fig6}(b). 
One can conclude that the new TB parameter set has lifted the valley degeneracy and produced multiple sub-plateaus illustrated by arrows in Fig.~\ref{fig6}(b).
Careful observers may notice seminal plateaus on the results calculated by 3NN$_I$ as well [Fig.~\ref{fig6}(a)]. Red-hallow circles show the conductance of the relevant two-terminal S-GWG (only one arm) with similar parameters and 3NN$_I$ energy set as a reference. 
Comparison between red-hallow circles, dashed and solid lines in Fig.~\ref{fig6}(a) tells us, the conductances of YJ-GWG almost follow the conductance of a relevant S-GWG. In Fig.~\ref{fig6}(b), it also shows 3NN$_{II}$ approximation results in a shift of conductances toward the positive side of energy axis. 
\vspace{-1em}
\subsection{Bandstructure Remarks}
\begin{figure}
  \begin{center}
	\includegraphics[width=8.4cm]{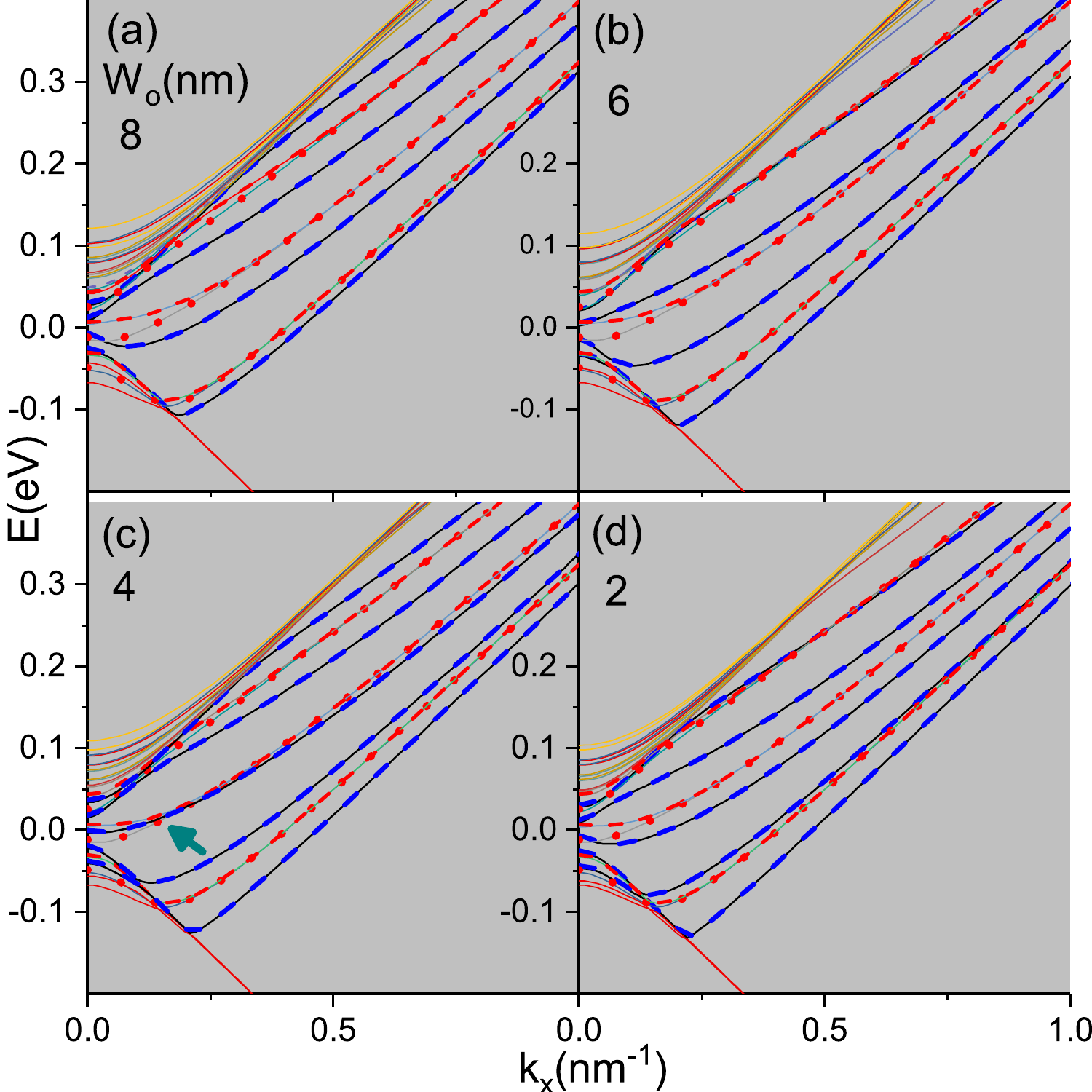}
   \end{center}
	\caption{\label{fig7} (Color plot) Quasi-one dimensional bandstructure of supercells that are extracted from the proposed YJ-GWG, Fig.~\ref{fig1}(b). A blue dashed and a solid line show two-fold degenerate bands of a AGNR's supercell at the left side of device in presence of single well, whereas a red dashed,a red dotted and two solid lines together show four-fold degenerate subbands of a AGNR supercell at the right side of YJ-GWG where two wells are separated by $W_{MB}$. 3NN$_I$ is employed.}
\end{figure}

In order to obtain more insight on our results, we have calculated quasi-one dimensional bandstructure for two supercells, one at the left end of AGNR, e.g., x = 10~nm (with single-well) and the other at the right end, e.g., x = 90~nm (with double well). The positions of those supercells are specified by blue and red dots in Fig.~\ref{fig1}(b). The calculated bands, which contain sixteen conduction bands and one valance band in the positive momentum space, are shown in Fig.~\ref{fig7}(a)-(d). 
Thus, it is possible  to track the movement of subbands in energy space at different $W_{O}$. A single-well possess two-fold degenerate bands which are highlighted by blue dashed and black narrow solid lines whereas double-well results in four-fold degenerate bands which are denoted by a red dashed, a red dotted line and two (multi-color) narrow solid lines together. 
Subbands belong to the barriers, which are located above the highlighted bands, are shown by multi-color narrow solid lines. All panels in Fig.~\ref{fig7} show that change in $W_{O}$ (or $W_{MB}$) does not shift the energy bands of double-well substantially, whereas similar change of $W_{O}$ shift down the bands of a single well. Note that decrease in $W_{O}$ leads to a similar increase in $W_{MB}$, because $W_O/2+W_{MB}/2=W_{be}$, where $W_{be}$ is fixed. Thus adjusting $W_{O}$ is a practical method to match energy channels at the two ends of scattering area. This diagram also explains why the smooth conductance occurs when $W_O=4$~nm at Fig~\ref{fig3}(c). We can see the two types of energy bands, shown by a tip of arrow, cross each other around the Dirac point at Fig.~\ref{fig7}(c). The calculated band structures indicate that lower energy bands with $E<0$ do not participate in the quantized transport.

\begin{figure}
	\begin{center}
		\includegraphics[width=8cm]{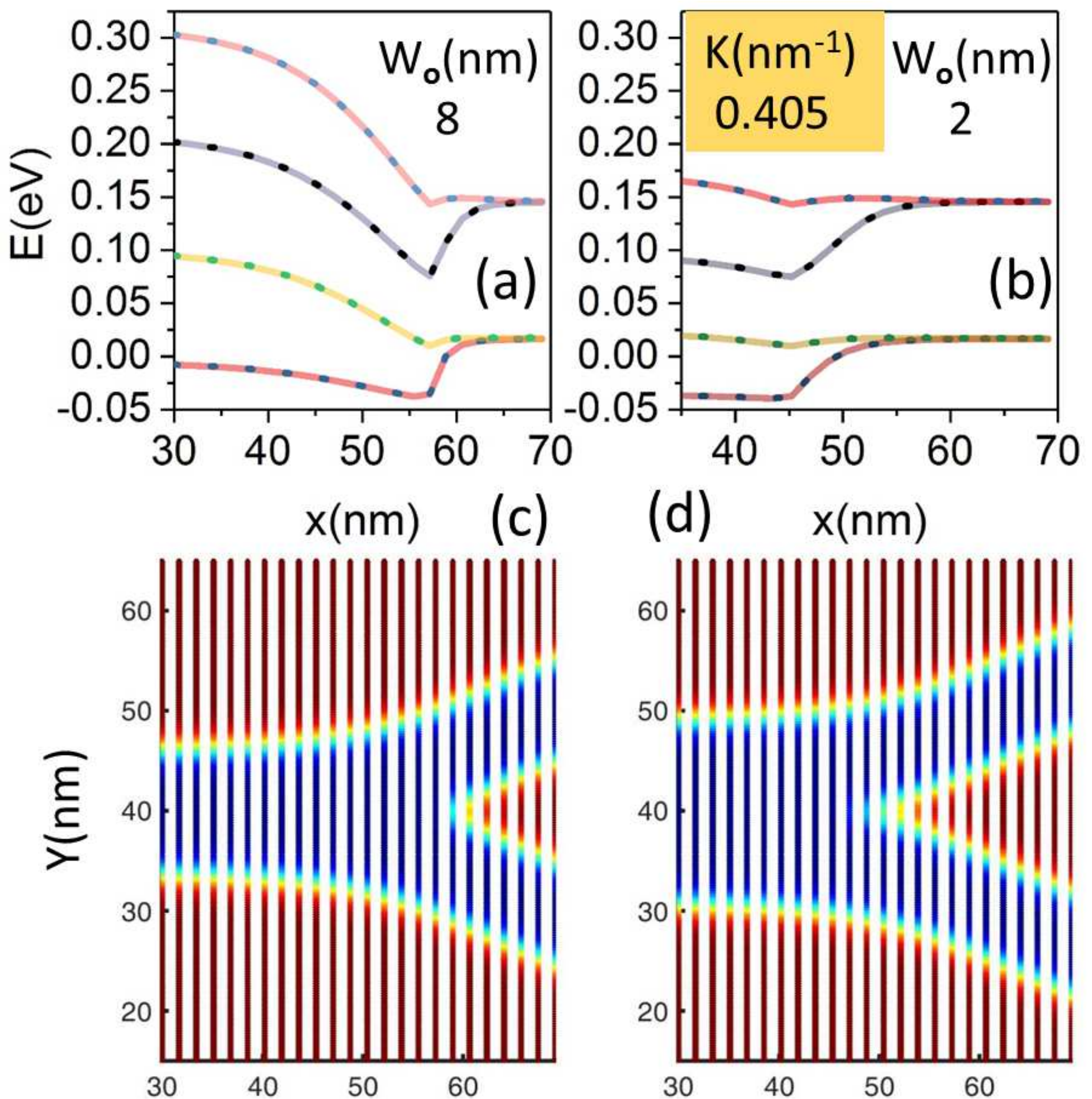}
	\end{center}
	\caption{\label{fig8} Spatial-resolved quasi-one dimensional energy spectrum for smooth YJ-GWG around splitting point. (a) $W_{o}$=8~nm. (b) $W_{o}$=2~nm. (c) On-site potential for selected supercells of (a). (d) On-site potential for selected supercells of (b).}
\end{figure}

\proofreadtrue
It is possible to explain a physical reason of our theoretical observation from the energy point of view. It is widely accepted that scattering happens as a result of mismatch between energy levels along the transport direction. For example, impurity levels might destructively couple to transport channels and smear the quantization of conductance. Here in YJ-GWG, one can expect minimum scattering if energy channels merge to each other with minimum misalignment around Y-junction. Therefore, we have further investigated our study by extracting spatial-resolved energy spectrum, i.e. waveguide energy channels, of few super-cells along x and around the splitting point, as plotted in Fig.~\ref{fig8}(a) and (b) for a fix momentum k$_x$= 0.405 ($nm^{-1}$). Relative on-site potential used for the calculation in Fig.8 (a) and (b) are plotted in Fig.~\ref{fig8}(c) and (d), respectively. One can see that two-fold degenerate energy bands (second and fourth levels) on the left side of Y-junction merge to four-fold degenerate bands on the right side. The amount of energy level variations are much less for the case of W$_o$ = 2nm as compare to W$_o$ = 8nm. Note that, one can expect similar merging trends for other higher momentums, since the E-k$_x$ relationship is linear for other larger values of momentum. From band structure studies, Fig.~\ref{fig8}(a) and (b), one can see obviously that the energy spacing between first and second four-fold bands, spatially available on the right side of device, is about 150 meV which is noticeably large and consistent with the width of first flat plateaus on the energy range at Fig.~\ref{fig3}.
\vspace{-1em}
\subsection{Discussion}
After the realization of quantized conductance on quantum point contact (QPC), Y-junction structures on modulation-doped heterostructures have been proposed and fabricated\cite{van1988quantized,maao1994quantum,shorubalko2003tunable}. Different terminologies have been used to address similar devices such as Y-branch switch (YBS) and three-terminal ballistic junction (TBJ). A gate defined YBS hardly works in ballistic regime or in another world the quantization of conductance is rarely reported as a prominent characteristic of device\cite{shorubalko2003tunable}. Whereas, peaks on the derivative of conductance on wet-etched version of YBS demonstrate the presence of quantization\cite{worschech2005nonlinear,jones2005quantum}. This is probably because subband-spacings in a gate defined 1D quantum wire are limited to the order of a few meV while larger energy spacing in a wet-etched GaAs/AlGaAs quantum wires, typically on the order of a few tens of meV, have been reported.  Note that, the conductance of wet-etched YBS does not exhibits flat plateaus and an exact identical I-V characteristic have not been reported.  On the other hand, carbon nanotube Y-junction (CNT-YJ), T-shaped GNR (T-GNR) and few Cross-shape GNR (C-GNR) have been widely studied theoretically\cite{andriotis2001rectification, andriotis2002transport}. The conductance of CNT-YJ shows oscillatory behavior in theoretical studies which merely depends on geometry of system. To the best our knowledge quantization of conductance does not report on chemically synthesized Y-junctions nanotube and both T-GNR and C-GNR do not exhibit quantization of conductance as well\cite{ouyang2009transport, bandaru2005novel, brandimarte2017tunable}.  A prominent benefit of the proposed structure is the flat plateaus on conductance of a smooth YJ-GWG, Fig.~\ref{fig3}(d), which is not reported in other possible carbon base splitters so far. The optimization of the width of incoming path on a smooth-bended YJ-GWG can be considered as a viable method to achieve the goal of coherent splitting. YJ-GWG can be a gateway for the realization of possible carbon-base electron interference devices, as well as interconnect in graphene-based spintronic devices.
\section{Conclusions}
In conclusion, we have exploited the possibility of coherent transport through a Y-junction graphene waveguide. Results show that if the incoming path of YJ-GWG is wide enough, the conductances of both paths are identical to the single graphene waveguide. It has been shown that smoothly bended GWG provide much smoother plateaus on the conductance. The primitive YJ-GWG enlarged by a scale factor of 1.5 and the results of transport study show that the quantization of conductance is well preserved in a larger device. In addition, we have shown that the influence of different TB energy sets on the calculation of conductance. A quasi-one dimensional bandstructure calculation was also performed to explain our observations.
\section*{Acknowledgments}
This work was supported by the National Key Research and Development Program of China (Grant No. 2016YFA0301700), the National Natural Science Foundation of China (Grants No. 11625419），and the Anhui Initiative in Quantum Information Technologies (Grants No. AHY080000). We should also thank Chinese Academy of Sciences and The World Academy of Science for the advancement of science in developing countries.

\appendix
\section{Y-junction on-site potential}
\begin{figure}
	\includegraphics[width=9cm]{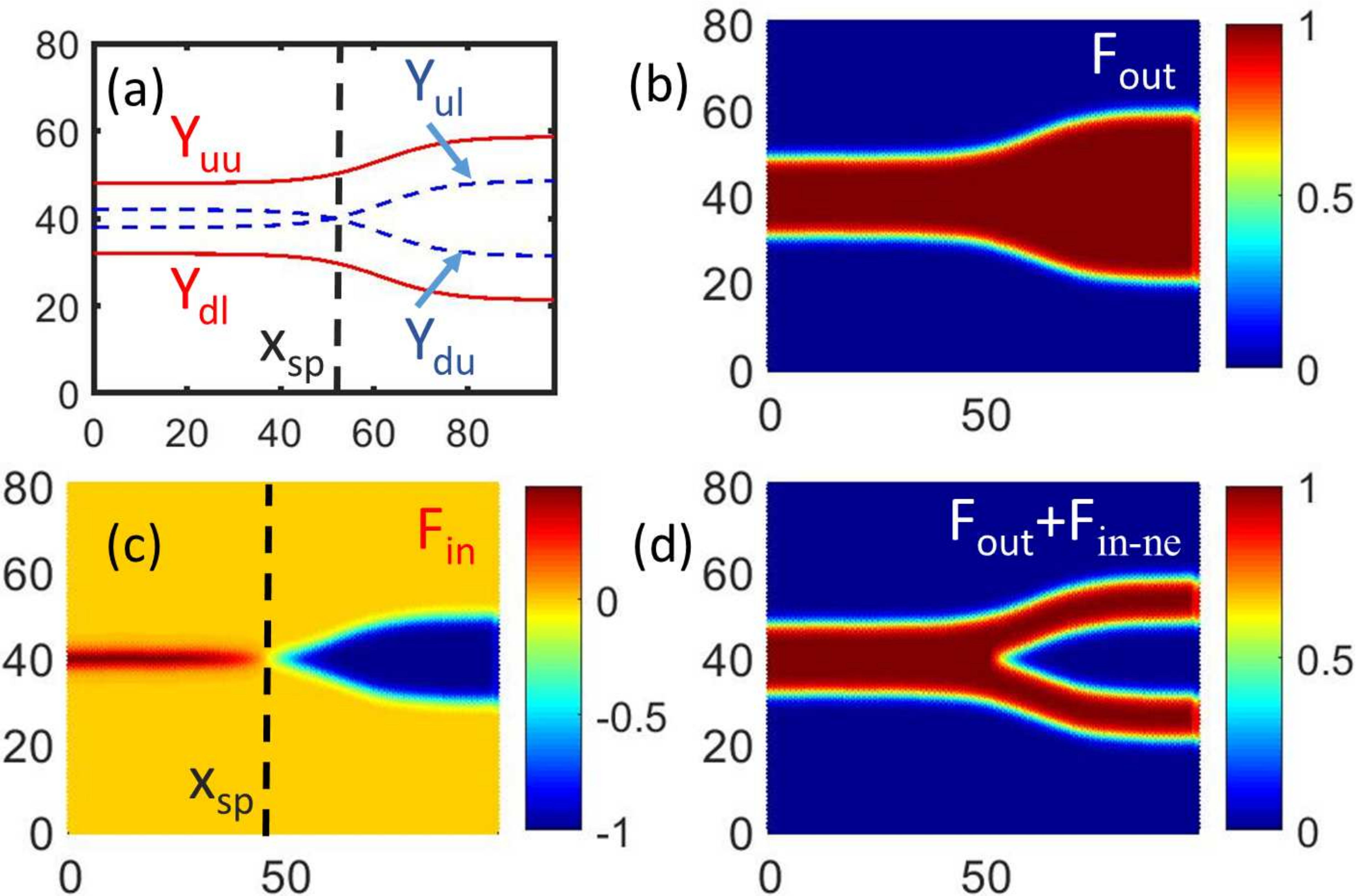}
	\caption{\label{fig9} (color online) Steps to build an on-site potential for YJ-GWG with smooth relaxation. (a) Border lines with which two bended S-GWG can be defined. (b) $F_{out}$. (c) $F_{in}$. Note $F_{in-ne}$ refers to $F_{in}$ on the right side of x$_{sp}$. (d) $F_{out}+F_{in-ne}$.}
\end{figure}

We have accomplished the following steps to build a 2D on-site potential of YJ-GWG including relaxation distance where values of potential energy inside and outside of Y-junction are u$_{in}$ and u$_{out}$.\\ 
Step 1~; Calculate functions, $Y_{du}(x_i)$ and $Y_{dl}(x_i)$, to define upper and lower edge lines of the downward bended S-GWG (i.e., downward lines in Fig.~\ref{fig9}(a)),
\begin{eqnarray}
Y_{du}(x_i)&=&(Y_0+W_O/2)+W_{be}(f(x_i)-1)
\\
Y_{dl}(x_i)&=&Y_{du}(x_i)-W_{da2}
\end{eqnarray}
and $Y_{ul}(x_i)$ and $Y_{uu}(x_i)$ to define lower and upper borders of the upward bended S-GWG (i.e., upward lines in Fig.~\ref{fig9}(a)) 
\begin{eqnarray}
Y_{ul}(x_i)&=&(Y_0-W_O/2)+W_{be}(1-f(x_i))
\\
Y_{uu}(x_i)&=&Y_{ul}(x_i)+W_{da1}
\end{eqnarray}
where $f(x_i)=1/(1+exp((x_i-x_0)/\sigma_x)$ is the Fermi function which is employed only to produce smooth variation along the x direction. $Y_0=W/2$ sets the center of YJ-GWG on the y direction,  $x_0$ is the inflection point of bended lines on the x direction and $\sigma_x$ defines the sharpness of bending in this direction. A bigger $\sigma_x$ provides a smooth and adiabatic variation along the x direction. Throughout all studies with smooth bending, the value of $\sigma_x$ is $L/14$. \\
Step 2~; Once again Fermi function is employed to establish another smooth step-like functions in y-direction which their inflection points define by making use of previously determined border functions i.e., $Y_{ul,uu,du,dl}(x_i)$ as follow;
\begin{eqnarray}
F_{bo}(x_i,y_i)=1/(1+exp((y_i-Y_{bo}(x_i))/\sigma_y)
\end{eqnarray}
where the index $bo$ refers to one of \textit{ul}, \textit{uu}, \textit{du} and \textit{dl}. We should remind that $x_i$ and $y_i$ are x and y coordinations of i-th atom. $\sigma_y$ determines relaxation distance ($\Delta W$) on transverse direction. One can easily conclude $\Delta W=2\sigma_yLn(9)\approx4.4\sigma_y$ if  $F_{bo}= 0.9$ and $F_{bo}= 0.1$ approximate upper and lower limits of F$_{bo}$. We have considered $\Delta W\approx $4 nm on all of our structures.\\
Step 3~; The subtraction $F_{out}=(F_{uu}-F_{dl})$ provide a function which smoothly varies from zero on both side-barrier areas to the one on the middle of outer lines i.e., $Y_{uu}$ and $Y_{dl}$, red solid lines in Fig.~\ref{fig9}(a). $F_{out}$ is plotted in Fig.~\ref{fig9}(b). Utilization of Fermi function on step 2 ensures that the full width of half maximum (FWHM) of waveguide is equal to the distance between border lines because the value of $F_{bo}$s are 1/2 on inflection points. Note that $Y_{du}$ and $Y_{ul}$, dashed lines in Fig.~\ref{fig9}(a), cross each other on the splitting point ($Y_0,x_{sp}$). At the same time, for $x_i>x_{sp}$ the subtraction $F_{in}=(F_{du}-F_{ul})$ provide us a function which smoothly varies from negative one on the middle-barrier (MB) region, the areas that is trapped between inner border lines (\textit{du} and \textit{ul}), to zero on side-barrier areas, see Fig.~\ref{fig9}(c). $F_{in}$ gives a positive value for overlapping area $x_i<x_{sp}$ which we cancel it by signing the positive values to zero and call the new function $F_{in-ne}$. 
\\
Step 4~; $F_{out}+F_{in-ne}$ delivers a Y-splitter function with 
zero value on all (three) barriers and positive one within the Y-branch, see Fig.~\ref{fig9}(d).\\
Step 5~; Finally, proper estimation of on-site potential of YJ-GWG is given by adding potential amplitudes as follow
\begin{eqnarray}
v_i(x_i,y_i)=u_{out}+(u_{in}-u_{out})[F_{out}+F_{in-ne}].
\end{eqnarray}
One can build a kink-shape on-site potential with similar process except kink-shape border lines must be taken into account instead of smooth lines in step 1.

%
\end{document}